\documentclass{aa}
\usepackage{graphicx}
\usepackage[varg]{txfonts}
\usepackage{cases}
\usepackage{natbib}

\begin{document}
   \title{Fast degradation of the circular flare ribbon on 2014 August 24}

   \author{Q. M. Zhang\inst{1,2}, S. H. Yang\inst{2,3}, T. Li\inst{2,3}, Y. J. Hou\inst{2,3}, and Y. Li\inst{1}}

   \institute{Key Laboratory of Dark Matter and Space Science, Purple Mountain Observatory, CAS, Nanjing 210033, PR China \\
              \email{zhangqm@pmo.ac.cn}
              \and
              CAS Key Laboratory of Solar Activity, National Astronomical Observatories, Chinese Academy of Sciences, Beijing 100101, PR China \\
              \and
              School of Astronomy and Space Science, University of Chinese Academy of Sciences, Beijing 100049, PR China \\
              }

   \date{Received; accepted}
    \titlerunning{Degradation of the circular flare ribbon}
    \authorrunning{Zhang et al.}

 \abstract
   {The separation and elongation motions of solar flare ribbons have extensively been investigated. The degradation and disappearance of ribbons have rarely been explored.}
   {In this paper, we report our multiwavelength observations of a C5.5 circular-ribbon flare associated with two jets (jet1 and jet2) on 2014 August 24, 
   focusing on the fast degradation of the outer circular ribbon (CR).}
   {The flare was observed in ultraviolet (UV) and extreme-ultraviolet (EUV) wavelengths by the Atmospheric Imaging Assembly (AIA) on board the Solar Dynamics Observatory (SDO) spacecraft.
   Soft X-ray (SXR) fluxes of the flare in 0.5$-$4 and 1$-$8 {\AA} were recorded by the GOES spacecraft.}
   {The flare, consisting of a short inner ribbon (IR) and outer CR, was triggered by the eruption of a minifilament.
   The brightness of IR and outer CR reached their maxima simultaneously at $\sim$04:58 UT in all AIA wavelengths.
   Subsequently, the short eastern part of CR faded out quickly in 1600 {\AA} but gradually in EUV wavelengths.
   The long western part of CR degraded in the counterclockwise direction and experienced a deceleration. 
   The degradation was distinctly divided into two phases: phase I with faster apparent speeds (58$-$69 km s$^{-1}$) and phase II with slower apparent speeds (29$-$35 km s$^{-1}$).
   The second phase stopped at $\sim$05:10 UT when the western CR totally disappeared.
   Besides the outward propagation of jet1, the jet spire experienced untwisting motion in the counterclockwise direction during 04:55$-$05:00 UT.}
   {We conclude that the event can be explained by the breakout jet model.
   The coherent brightenings of the IR and CR at $\sim$04:58 UT may result from the impulsive interchange reconnection near the null point, 
   whereas sub-Alfv\'{e}nic slipping motion of the western CR in the counterclockwise direction indicates the occurrence of slipping magnetic reconnection.
   Another possible explanation of the quick disappearance of the hot loops connecting to the western CR is that they are simply reconnected sequentially without
   the need for significant slippage after the null point reconnection.}


 \maketitle

\section{Introduction} \label{s-intro}
Solar flares are one of the most powerful activities in the solar atmosphere \citep{fle11}. Amount to 10$^{29}-$10$^{32}$ erg magnetic free energy is impulsively released 
via magnetic reconnection within tens of minutes to a few hours \citep{ems12,yang15,li17a,chen19b,hong19,zqm19b}. 
In the context of collisional thick-target flare model, high-energy electrons stream downward into the cool and dense chromosphere where electrons are stopped 
by Coulomb collisions and lose energy, heating the localized plasma to $\sim$10$^7$ K and driving chromospheric evaporation \citep{bro71,fis85}. 
Emissions at the flare ribbons or kernels increase dramatically in optical, 
ultraviolet (UV), extreme-ultraviolet (EUV), hard X-ray (HXR), and microwave wavelengths \citep[e.g.,][]{kru11,fle13,sha14,jing16,chen19a}. 
After reaching the peak value, the brightness of ribbons begins to decrease as a result of conductive and radiative cooling \citep{car95}.

As magnetic reconnection in the corona proceeds, the double flare ribbons separate with time \citep{qiu02,qiu04}. 
The rate of reconnection could be estimated based on the separation speed (20$-$100 km s$^{-1}$) and line-of-sight (LOS) magnetic field strength on the conjugate ribbons. 
High-cadence and high-resolution observations 
reveal that the distance between the double centroids of H$\alpha$ bright kernels decreases at the rising phase of the HXR spikes and increases after the flare peak time \citep{ji04,ji06}.
Apart from the separation, the brightening may propagate along the flare ribbons at speeds of a few to 100 km s$^{-1}$ \citep{qiu09,qiu17}. 
During three-dimensional (3D) magnetic reconnection within the thin quasi-separatrix layer \citep[QSL;][]{dem96} where the gradient in field-line linkage 
is sharp and electric current accumulates, the flare ribbons propagate along the intersection of QSL with the photosphere \citep[e.g.,][]{aul07,jan13,li15,jan16,sav16,zhao16}.

Circular-ribbon flares (CRFs) are a special kind of flares whose ribbons have circular or quasi-circular shapes \citep[e.g.,][]{mas09,wang12,her17,li18,chen19a,hou19}.
In most cases, they are triggered by filament eruptions \citep{yang18,li19}.
The 3D magnetic configuration associated with a CRF usually features a magnetic null point and the corresponding fan-spine skeleton \citep{xu17,li19}.
The inner ribbon (IR) is shorter and brighter than the outer circular ribbon (CR), implying that the energy deposition is more concentrated in the inner ribbon \citep{zqm16a,zqm16b}.
The chromospheric brightening of a CRF is not always simultaneous but sequential, with apparent speeds of ten to 220 km s$^{-1}$ \citep{wang12,li17b,rom17,xu17,li18}.
So far, the degradation and disappearance of flare ribbons have rarely been investigated.

In this paper, we report our multiwavelength observations of the C5.5 CRF on 2014 August 24, focusing on the fast degradation of the outer CR.
In Sect.~\ref{s-data}, we describe the data analysis. Results are presented in Sect.~\ref{s-res}.
We compare our findings with previous models and give a brief summary in Sect.~\ref{s-disc}.

\section{Observations and data analysis} \label{s-data}
The C5.5 flare in NOAA active region (AR) 12149 (N10E44) was observed by the Atmospheric Imaging Assembly \citep[AIA;][]{lem12} on board the Solar Dynamics Observatory (SDO).
AIA takes full-disk images in two UV (1600 and 1700 {\AA}) and seven EUV (94, 131, 171, 193, 211, 304, and 335 {\AA}) wavelengths.
The level\_1 data were calibrated using the standard solar software program \texttt{aia\_prep.pro}.
Soft X-ray (SXR) light curves of the flare in 0.5$-$4 and 1$-$8 {\AA} were recorded by the GOES spacecraft.
The observational parameters are listed in Table~\ref{tab-1}.

\begin{table}
\centering
\caption{Description of the observational parameters.}
\label{tab-1}
\begin{tabular}{ccccc}
\hline\hline
Instrument & $\lambda$   & Time & Cad. & Pix. Size \\ 
                  & ({\AA})         &  (UT) & (s)           & (\arcsec) \\
\hline
SDO/AIA & 94$-$335 & 04:30$-$06:00 & 12 & 0.6 \\
SDO/AIA & 1600        & 04:30$-$06:00 & 24 & 0.6 \\
GOES     & 0.5$-$4.0 & 04:30$-$06:00 & 2.05 & ... \\
GOES     & 1$-$8    & 04:30$-$06:00 & 2.05 & ... \\
\hline
\end{tabular}
\end{table}

\section{Results} \label{s-res}
\begin{figure}
\includegraphics[width=8cm,clip=]{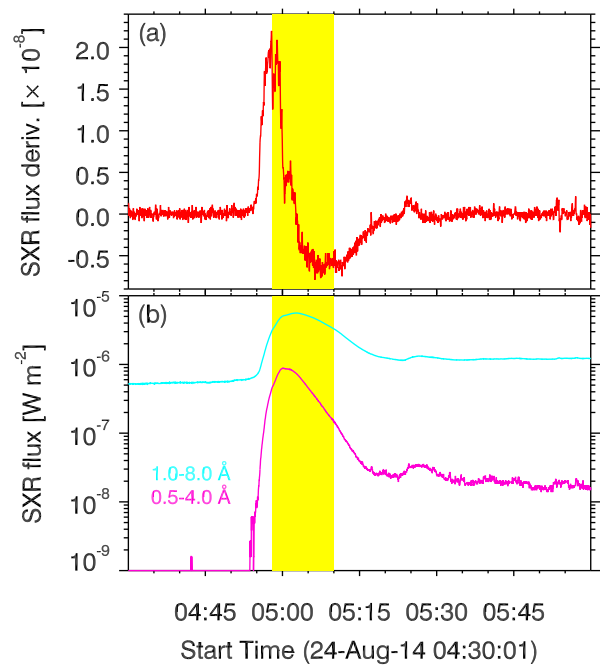}
\centering
\caption{Bottom panel: SXR light curves of the C5.5 flare in 0.5$-$4 {\AA} (magenta line) and 1$-$8 {\AA} (cyan line).
Top panel: Time derivative of the 1$-$8 {\AA} flux. In each panel, the yellow region stands for the time of fast degradation of the outer circular ribbon.}
\label{fig1}
\end{figure}

In Fig.~\ref{fig1}, the bottom panel shows the SXR light curves of the flare. It is clear that the SXR emissions started to rise at $\sim$04:55 UT and 
reached the peak values at $\sim$05:02 UT, which were followed by a gradual decay phase until $\sim$05:25 UT. 
Since the HXR flux during the flare was not available, we take time derivative of the 1$-$8 {\AA} flux as a HXR proxy according to the Neupert effect.
In Fig.~\ref{fig1}(a), two spikes at 04:58:36 UT and 04:59:18 UT are remarkable, which are concurrent with the primary and second peaks in 1600 {\AA} (see Fig. 1(d) in \citet{zqm19a}).
A plausible reason for these spikes is intermittent magnetic reconnection and subsequent energy input in the chromosphere by nonthermal electrons \citep{zqm16b}.

\begin{figure}
\includegraphics[width=9cm,clip=]{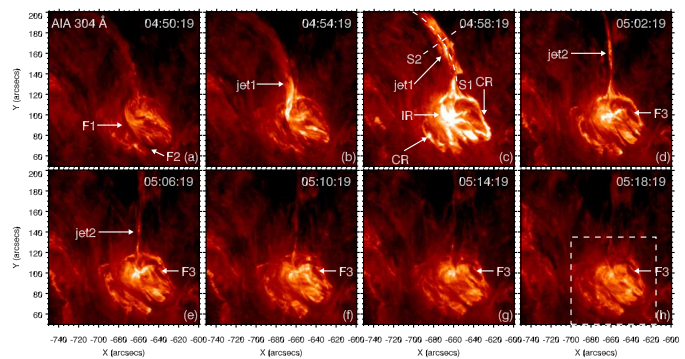}
\centering
\caption{Eight snapshots of the AIA 304 {\AA} images during 04:50$-$05:18 UT. 
The white arrows point to the circular ribbon (CR), inner ribbon (IR), two jets (jet1, jet2), and three filaments (F1, F2, F3).
Two slices (S1 and S2) in panel (c) are used for investigating the propagation and rotation of jet1, respectively.
The white dashed box in panel (h) signifies the field of view of Fig.~\ref{fig3}.
Evolution of the flare is shown in a movie (\textit{anim304.mov}) available online.}
\label{fig2}
\end{figure}

In Fig.~\ref{fig2}, eight snapshots of the AIA 304 {\AA} images illustrate the evolution of flare (see also the online animation \textit{anim304.mov}). 
Before the flare, there were three minifilaments (F1, F2, and F3) at the flaring region.
Impulsive eruption of the eastern filament (F1) resulted in strong heating of the filament and generation of the first blowout jet (jet1) (see panels (b)-(c)).
Meanwhile, magnetic reconnection at the null point may be triggered \citep{pri09,zqm12,bau13}. The accelerated nonthermal electrons propagate downward 
along the fan-spine field lines and precipitate in the chromosphere. The impulsive heating results in the CRF, which consists of a compact IR and the surrounding CR.
The brightness of IR and CR reached their maxima at $\sim$04:58 UT. Shortly afterwards, another jet (jet2) appeared adjacent to jet1 and propagated northward (see panel (d)). 
Subarcsecond blobs (plasmoids) with an average size of 0{\farcs}78 were detected in the jets \citep{zn19}.
The southern minifilament (F2) and western minifilament (F3) were undisturbed and remained there (see panels (d)-(h)).

\begin{figure}
\includegraphics[width=8cm,clip=]{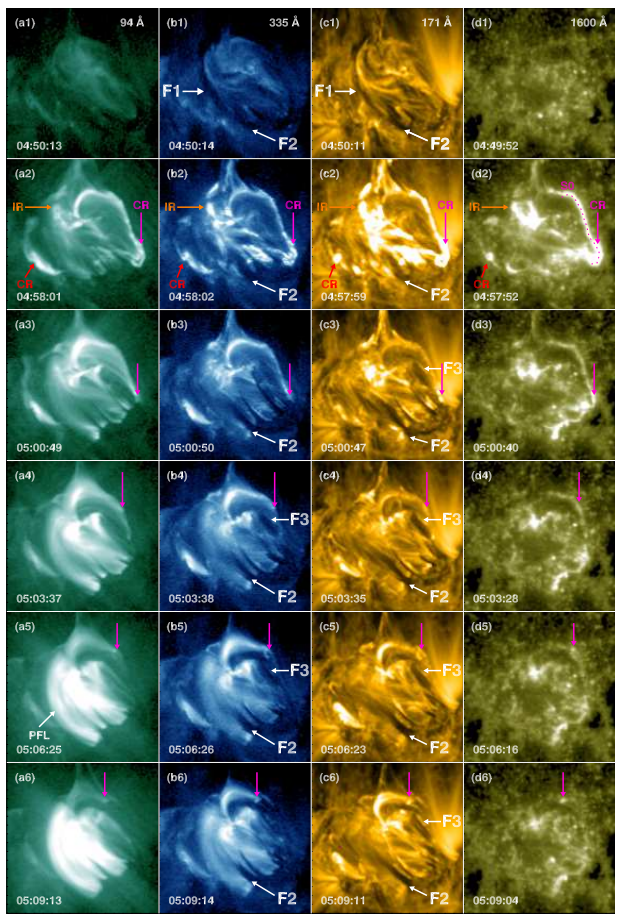}
\centering
\caption{Evolution of the flare ribbons observed in AIA 94 {\AA} (a1-a6), 335 {\AA} (b1-b6), 171 {\AA} (c1-c6), and 1600 {\AA} (d1-d6).
The white arrows point to F1, F2, F3, and PFL. The orange arrows point to the IR. The red and magenta arrows point to the eastern and western parts of CR, respectively.
Evolution of the CR is shown in a movie (\textit{animrb.mov}) available online.}
\label{fig3}
\end{figure}

In Fig.~\ref{fig3}, the left column (a1-a6) shows the evolution of flare ribbons in 94 {\AA} ($\log T\approx6.8$).
It is seen that bright post-flare loop (PFL) became evident as magnetic reconnection proceeded.
During 04:58$-$05:04 UT, converging motion of hot plasma along the PFL towards the loop top was detected, which is a clear indication of chromospheric evaporation \citep{zqm19a}.
The second, third, and fourth columns of Fig.~\ref{fig3} show the evolution of ribbons in 335 {\AA} ($\log T\approx6.4$), 171 {\AA} ($\log T\approx5.8$), and 1600 {\AA}, respectively.
As mentioned above, the brightness of IR and outer CR reached their maxima simultaneously at $\sim$04:58 UT in all AIA wavelengths, which is consistent with the HXR peak time 
(see Fig.~\ref{fig1}(a)). Subsequently, the short eastern part of CR, which is pointed by the red arrows, faded out quickly in 1600 {\AA} (see panels (d2)-(d6)) but gradually in EUV wavelengths.
The long western part of CR, which is pointed by the magenta arrows, did not fade out coherently as expected. Instead, the CR degraded in the counterclockwise direction during 04:58$-$05:10 UT.

\begin{figure}
\includegraphics[width=9cm,clip=]{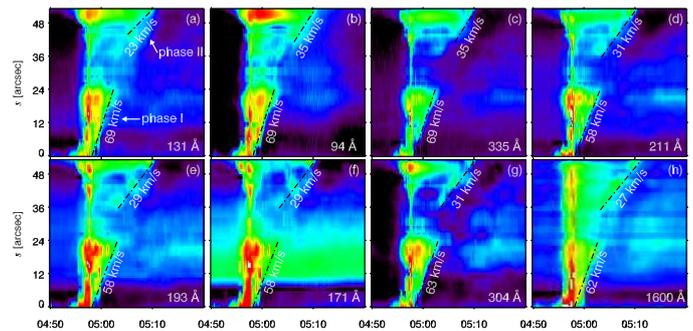}
\centering
\caption{Time-distance diagrams of S0 in various wavelengths, showing the fast degradation of the western CR during 04:58$-$05:10 UT.
The apparent speeds are labeled. $s=0$ and $s=52\farcs8$ on the $y$-axis denote the southern and northern endpoints of S0, respectively.}
\label{fig4}
\end{figure}

In Fig.~\ref{fig3}(d2), a curved artificial slice (S0) along the western CR is selected. The time-distance diagrams of S0 in various wavelengths are displayed in Fig.~\ref{fig4}.
It is obvious that the CR brightened simultaneously at $\sim$04:58 UT in all wavelengths. Then, the CR degraded counterclockwise and experienced a deceleration. 
The degradation was obviously divided into two phases: phase I with faster apparent speeds (58$-$69 km s$^{-1}$) and phase II with slower apparent speeds (29$-$35 km s$^{-1}$).
The average speed of phase I is $\sim$2.1 times higher than that of phase II. The second phase stopped at $\sim$05:10 UT when the western CR totally disappeared.

\begin{table}
\centering
\caption{Apparent speeds (km s$^{-1}$) of degradation of the western CR in different AIA wavelengths ({\AA}).}
\label{tab-2}
\begin{tabular}{c|cccccccc}
\hline\hline
 $\lambda$ & 131 & 94 & 335 & 211 &  193 & 171 & 304 & 1600 \\
\hline
ph\_I & 69 & 69 & 69 & 58 & 58 & 58 & 63 & 62 \\
ph\_II & 23 & 35 & 35 & 31 & 29 & 29 & 31 & 27 \\
\hline
\end{tabular}
\end{table}

\section{Discussion and summary} \label{s-disc}
To explain how accelerated particles at a reconnection site could finally propagate along the Earth-connected open flux tube, \citet{mas12} performed a 3D numerical simulation
under the magnetic topology of an asymmetric coronal null point, with a closed fan surface and an open outer spine. It is found that interchange magnetic reconnection takes place 
when field lines anchored below the fan dome reconnect at the null point and jump to the open magnetic domain. The reconnected field lines near the outer spine keep reconnecting
continuously and show apparent slipping motion, which is quite different from the case in 2D.
Recently, \citet{wyp18} performed 3D numerical simulations of coronal jets with filaments, confirming the breakout jet generation mechanism \citep{wyp17}. 
The untwisting, blowout jets are produced by a combination of an untwisting flux rope and plasma accelerated by the flare reconnection below. 
\citet{doy19} studied a helical jet associated with a C1.5 flare triggered by a confined filament eruption on 2013 June 30. 
The erupting filament material was partially transferred into the outflow jet along large-scale, overlying coronal loops. 
Additionally, the authors conducted a 3D MHD numerical simulation of a breakout jet in a closed-field configuration and found excellent qualitative agreement with observations.
Hence, it is concluded that the confined eruption with a rotating jet could well be explained by the breakout model \citep{wyp17,wyp18}.

\begin{figure}
\includegraphics[width=9cm,clip=]{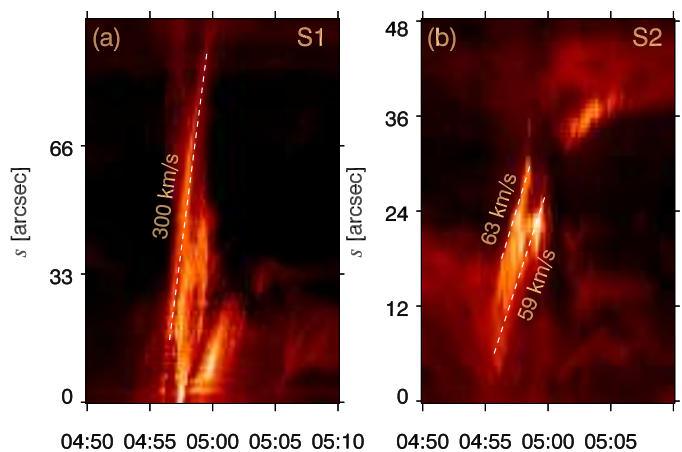}
\centering
\caption{Time-distance diagrams of S1 and S2 in 304 {\AA}. The apparent speeds are labeled. 
In panel (a), $s=0$ and $s=99\arcsec$ on the $y$-axis denote the southern and northern endpoints of S1.
In panel (b), $s=0$ and $s=48\arcsec$ on the $y$-axis denote the eastern and western endpoints of S2.}
\label{fig5}
\end{figure}

In our study, the first jet (jet1) associated with the C5.5 CRF also shows untwisting motion. In Fig.~\ref{fig2}(c), two artificial slices (S1 and S2) along and across the jet spire are selected. 
The time-distance diagrams of S1 and S2 in 304 {\AA} are displayed in the left and right panels of Fig.~\ref{fig5}, respectively. 
The apparent propagation speed of jet1 during 04:55$-$05:00 UT is calculated to be $\sim$300 km s$^{-1}$.
In panel (b), it is obvious that the jet spire shows untwisting motion in the counterclockwise direction when looking down and splits into two threads during 04:55$-$05:00 UT.
The apparent speeds of untwisting motion are estimated to be $\sim$59 and $\sim$63 km s$^{-1}$. It is noted that jet2 does not show significant rotation, possibly due to that the twist (helicity) 
in the erupting filament (F1) has entirely been transported to the large-scale coronal loops during jet1. Therefore, the CRF associated with an untwisting jet triggered by the eruption of
a minifilament can be interpreted by the breakout model \citep{wyp18}. The coherent brightenings of the IR and CR at $\sim$04:58 UT may result from the impulsive interchange reconnection
near the null point, whereas sub-Alfv\'{e}nic slipping motion of the western CR in the counterclockwise direction is indicative of slipping magnetic reconnection. 
The apparent speeds of degradation are close to the speeds of elongation of the CR during the second episode in the event of \citet{li18}.
The deceleration of degradation from phase I to phase II suggests that magnetic reconnection tapered off.

It is revealed in the breakout jet model that the magnetic flux of strapping field overlying the erupting filament is transferred to the other side of the fan dome,
allowing the filament to reach the null point, where it is reconnected and produces a helical jet. Meanwhile, a flare current sheet is formed beneath the filament, 
so that the reconnected overlying flux is reconnected back to where the filament originated.
In Fig.~\ref{fig3}(a3), hot newly reconnected loops appeared, connecting to the western CR. 
Those hot loops might be the breakout loops on the other side of F1 (see also Fig. 9(b) in \citet{doy19}).
The sequential disappearance of those hot loops during 04:58$-$05:04 UT is not only concurrent with the fast degradation of western CR in phase I, 
but also consistent with the brightening of PFL to the east side of anemone (see Fig.~\ref{fig3}(a4-a6)).
The evolution is evidently illustrated in Fig. 9 of \citet{doy19} and is nicely in agreement with the breakout jet model by \citet{wyp18}.
Therefore, another possible explanation of the quick disappearance of the hot loops connecting to the western CR is that they are simply reconnected sequentially without
the need for significant slippage after the null point reconnection.

In this work, we report our multiwavelength observations of a C5.5 CRF associated with two jets on 2014 August 24.
The flare, consisting of a short IR and outer CR, was triggered by the eruption of a minifilament.
The brightness of IR and CR reached their maxima simultaneously at $\sim$04:58 UT in all AIA wavelengths.
Afterwards, the short eastern part of CR faded out quickly in 1600 {\AA} but gradually in EUV wavelengths.
The long western part of CR degraded in the counterclockwise direction and experienced a deceleration. 
The degradation was distinctly divided into two phases: phase I with faster apparent speeds (58$-$69 km s$^{-1}$) and phase II with slower apparent speeds (29$-$35 km s$^{-1}$).
The second phase stopped at $\sim$05:10 UT when the western CR completely disappeared.
Apart from the longitudinal propagation of jet1, the jet spire experienced untwisting motion in the counterclockwise direction during 04:55$-$05:00 UT.
We conclude that the analysis of this event is in favor of the breakout jet model.
The coherent brightenings of the IR and CR at $\sim$04:58 UT may result from the impulsive interchange reconnection near the null point, 
whereas sub-Alfv\'{e}nic slipping motion of the western CR in the counterclockwise direction indicates the occurrence of slipping magnetic reconnection.

\begin{acknowledgements}
The authors appreciate the referee for valuable suggestions and comments. We also thank K. Yang for discussions.
SDO is a mission of NASA\rq{}s Living With a Star Program. AIA data are courtesy of the NASA/SDO science teams.
This work is funded by NSFC grants (No. 11773079, 11790302, 11673035, 11773039, 11903050, 11790304, 11873095), 
the International Cooperation and Interchange Program (11961131002), the Youth Innovation Promotion Association CAS, 
the Science and Technology Development Fund of Macau (275/2017/A), 
the Strategic Priority Research Program on Space Science, CAS (XDA15052200, XDA15320301), 
and the project supported by the Specialized Research Fund for State Key Laboratories.
\end{acknowledgements}

\end{document}